\documentclass[journal,onecolumn]{IEEEtran}

\usepackage{cite}      
\usepackage[dvips]{graphicx} 
\usepackage{subfigure} 
\usepackage{url}        
\usepackage[intlimits]{amsmath}
\usepackage{amssymb}   
\usepackage{array}

\newcommand{\vect}[1]{\boldsymbol{#1}}
\newcommand{\E}{\mathbb{E}}
\newcommand{\argmax}{\operatornamewithlimits{arg\,max}}
\newcommand{\SNR}{\rho}
\def\euler{\gamma}
\def\d{\mathrm{d}}
\newcommand{\ol}[1]{\overline{#1}}
\newcommand{\ligning}[1]{\begin{align*} #1 \end{align*}}
\newcommand{\g}[1]{\Gamma_\textsc{#1}}
\newcommand{\G}[1]{\ol{\Gamma}_\textsc{#1}}

\newtheorem{proposition}{Proposition}

\begin{document}

\title{Multicell Zero-Forcing and User Scheduling on the Downlink of a Linear Cell Array}
\author{Hans~J{\o}rgen~Bang, David Gesbert.
\thanks{This work was presented in part at SPAWC'2009, Perugia, June 2009.}}
\maketitle

\begin{abstract}
Coordinated base station (BS) transmission has attracted much interest for its potential to increase the capacity of wireless networks. Yet at the same time, the achievable sum-rate with single-cell processing (SCP) scales optimally with the number of users under Rayleigh fading conditions. One may therefore ask if the value of BS coordination is limited in the many-user regime from a sum-rate perspective. With this in mind we consider multicell zero-forcing beamforming (ZFBF) on the downlink of a linear cell-array. We first identify the beamforming weights and the optimal scheduling policy under a per-base power constraint. We then compare the number of users $m$ and $n$ required per-cell to achieve the same mean SINR, after optimal scheduling, with SCP and ZFBF respectively. Specifically, we show that the ratio $m/n$ grows logarithmically with $n$. Finally, we demonstrate that the gain in sum-rate between ZFBF and SCP is significant for all practical values of number of users.
\end{abstract}


\section{Introduction}
In conventional cellular systems signal transmission and reception are done independently on a per-cell basis. This results in considerable inter-cell interference which ultimately limits the capacity.
However, by interconnecting the BSs and coordinate their actions the inter-cell interference can be greatly reduced~\cite{Karakayali2006,Dai2004}. 
A key driver for practical deployment of BS coordination is that the main complexity burden is on the network side and not the mobile users.

Recently there has been much work on the information theoretic nature of coordinated networks~\cite{Somekh2007,Weingarten2006}. In particular the downlink can be viewed as vector broadcast channel in which dirty paper coding (DPC) is the capacity achieving strategy. Unfortunately, for most practical applications DPC is prohibitively complex. Sub-optimal techniques with lower complexities such as linear precoding are therefore of great interest.
 
In this paper we consider multicell zero-forcing beamforming (ZFBF) together with multiuser scheduling. We are particularly keen to compare the resulting sum-rate per cell, with that of single-cell processing (SCP) and optimal scheduling. The reason for this is twofold. First of all, there is an inevitably increase in complexity with any BS coordination scheme relative to conventional SCP. To justify the use of BS coordination there must therefore be an accompanied gain in performance.
Second, under standard fading assumptions arbitrarily high sum-rates can be achieved with SCP by admitting sufficiently many users into the system. Furthermore, the asymptotic rate of increase with the number of users has shown to be optimal~\cite{Gesbert2007,Sharif2005}. 
A corollary to this is that there is little need for BS coordination with asymptotically many users. The practical implications of this result for the many but pre-asymptotic user regime is therefore of interest.

For analytical tractability we adopt a particularly simple network and interference model. Specifically, we assume a linear cell-array, where each user only receives a signal from the two closest BSs.  This is a slight modification of Wyner's classical model introduced in~\cite{Wyner1994}. For symmetry reasons we consider an infinite number of cells. 
However,  the alternative choice of a finite number of cells would have no qualitative impact on the results.

Importantly, we assume a per-base power constraint since the antennas are not co-located. The alternative choice of an overall power constraint in less realistic, but usually more attractive from an analytical point of view~\cite{Yu2005}. Fortunately, we will see that a per-base power constraint is easily tackled for the system model at hand.  Another key  assumption of this work is that full channel knowledge is available at the transmitter side. This is clearly hard to accomplish in a practical setting. BS coordination with reduced channel information is therefore an important topic~\cite{Boccardi2007,Papadogiannis2008}. However, we will not focus on this here. 

Similar network and interference models were recently used in \cite{Somekh2007} and \cite{Jing2008}, with the exception that the cells were arranged on a circular array. However, this difference is insignificant as the number of cells goes to infinity. In \cite{Somekh2007} the focus was on upper and lower bounds for the per-cell sum-rate under DPC. In particular, the per-cell sum-rate was shown to scale as $\log\log n$ with the number of users $n$ per cell. In \cite{Jing2008} the performances of several suboptimal network coordination strategies were characterized. However, no explicit expressions for ZFBF together with Rayleigh fading were given. In \cite{Somekh2009} ZFBF and multiuser scheduling were studied using a model where each user could see the three closest BSs. A suboptimal scheduling strategy was proposed and shown to scale optimally with the number of users. However, optimal scaling can also be achieved with SCP and is therefore not sufficient to justify ZFBF in itself.

\begin{figure*}
\centering
\centering
\includegraphics[width=.7\linewidth]{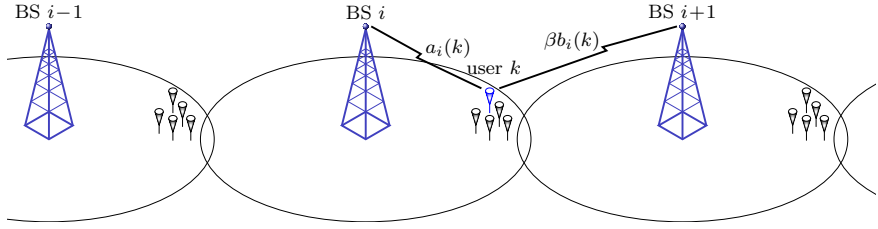}
\caption{Part of infinite linear cell-array. Each user receives a signal from the two closest BSs.}
\end{figure*}

The goal of this work is to to evaluate the benefit of multicell ZBBF over SCP in the many-user regime. To this end we  derive explicit expressions for a set of beamforming weights satisfying the zero-forcing criterion and a per-base power constraint. Based on this preliminary result we identify the optimal scheduling policy. To make a first comparison with SCP we note that the post-scheduling signal-to-interference-plus-noise ratio (SINR) can be viewed as the maximum of a random sample of size $n$. 
This observation allows us to draw on 
Extreme Value Theory (EVT)~\cite{Galambos1987,deHaan2006} 
to characterize the asymptotic behavior of the mean SINR with the number of users. 
We scrutinize our findings further by giving some exact result as well as several upper and lower bounds.
Notably, we derive asymptotic expressions for the number of users $m$ and $n$ required to attain the same mean SINR with SCP and ZFBF respectively. Put differently, we find the extra number users needed per cell to compensate for the lack of coordination with SCP. Interestingly, the ratio $m/n$ is not bounded, but grows logarithmically with the number of users $n$. 
Finally, we demonstrate that the difference in sum-rate between ZFBF and SCP is significant for all practical values of number of users.

\section{System Model}
We consider an infinite linear cell-array with $n$ users in each cell. We assume intra-cell TDM with synchronous time slots (scheduling intervals) across the network. The time slots are assumed to be sufficiently short for the channel coefficients to be constant within a slot, yet contain enough symbols to employ capacity achieving codes. In the following we will focus on an arbitrary symbol transmission interval within an arbitrary time slot and omit explicit reference to time. The received signal for user $k$ in cell $i$ is given by 
\begin{equation}
y_i(k)=a_{i}(k)x_i+\beta b_{i}(k)x_{i+1}+z_i(k),
\label{eq:channel_model}
\end{equation}
where $x_i$,$x_{i+1}$$\in \mathbb{C}$ are the antennae outputs from BS $i$ and BS $i\!+\!1$, $a_i(k)$,$b_i(k)$$\in \mathbb{C}N(0,1)$ are the corresponding fading coefficients and $z_i(k)$$\in \mathbb{C}N(0,1)$ is normalized Gaussian noise. 
The constant $\beta$$\in[0,1]$ reflects a difference in the path loss on the two signal paths.

In each time slot there is one user, denoted $k_i^*$, that is scheduled in each cell $i$. If we focus on the scheduled users we have the following input-output relationship
\begin{displaymath}
\vect{y}=H\vect{x}+\vect{z}	
\end{displaymath}
where $\vect{y}=\{y_i(k_i^*)\}_{i\in\mathbb{Z}}$, $\vect{x}=\{x_i\}_{i\in\mathbb{Z}}$, $\vect{z}=\{z_i(k_i^*)\}_{i\in\mathbb{Z}}$ are infinite column vectors and $H$ is a bidiagonal infinite matrix with
\begin{displaymath}
	[H]_{i,j}=\begin{cases}
	a_i(k_i^*), & i=j\\
	\beta b_i(k_i^*), &  i=j-1\\
	0      & \text{otherwise}.
		\end{cases}
\end{displaymath}
In the case of multicell linear beamforming (preprocessing) one applies a matrix $B$ such that $\vect{x}=B\vect{s}$  where $\vect{s}=\{s_i\}_{i\in\mathbb{Z}}$ is an infinite column vector. Here $s_i$ is the information symbol intended for user $k^*_i$. In order to fulfill a per BS power constraint we require $\E|x_i|^2\leq \SNR$. With the assumption $\E|s_i|^2=1$ this is equivalent to the $\ell^2$-norm of each row of B being no more than $\sqrt{\SNR}$.

Finally, full channel information is available to the BSs, while the users are aware of their own channel realizations and employ conventional single user receivers. 

\section{Single-Cell network bound}
As a reference we first consider the case with no inter-cell interference ($\beta=0$).  The channel model now reduces to
\begin{equation}
	y_i(k)=a_i(k)x_i+z_i(k).
\label{eq:scn_model}
\end{equation}
Conceptually this is equivalent to a network with one single isolated cell. The channel model in \eqref{eq:scn_model} is the prototype model for illustrating the potential gains of multiuser scheduling. The optimal scheduling policy is to select the user $k$ with the largest gain $|a_i(k)|$ in cell $i$ which yields the instantaneous SINR
\begin{displaymath}
	\Gamma^i_{\textsc{scn}}(n)=\max_{1\leq k\leq n}\SNR |a_i(k)|^2.
\end{displaymath}
In the sequel we will drop the index $i$ when denoting $\g{scn}^i(n)$ since its distribution is independent of the particular cell. To find the distribution of $\g{scn}(n)$ we first note that $\g{scn}:=\g{scn}(1)$ is exponentially distributed with cdf  
$$F_\textsc{scn}(x)=1-e^{-x/\SNR},\quad x\geq 0.$$
Since $\g{scn}(n)$ can be phrased as the largest order statistics of $\g{scn}$ the cdf $F_\textsc{scn}^n$ of $\g{scn}(n)$ is~\cite{David2003}
$$F_\textsc{scn}^n(x)=\bigr(1-e^{-x/\SNR}\bigl)^n,\quad x\geq 0.$$
It is well know that the corresponding mean is 
$$\E\,\g{scn}(n)=\int\nolimits_0^\infty x\,\d F_\textsc{scn}^n =\SNR H_n,$$
where $H_n:=\sum_{k=1}^n 1/k$ is the $n$th harmonic number~\cite{David2003}. The above expression can also be extended formally to all $y\in\mathbb{R}_+$ by using the analytical continuation of $H_n$,
$$H_y=\varphi(y+1)+\euler,$$
where $\varphi(\cdot)$ is the digamma function and $\euler=0.577..$ is the Euler constant~\cite{Gradshteyn}.


In the next sections we will demonstrate that the single-cell network (SCN) scenario upper bounds the performance of both SCP and ZFBF in a multi-cell network. However, it is worth pointing out that the SCN bound can be achieved in a multi-cell network by the use of DPC.

\section{Single-cell processing}
In conventional SCP networks all signal transmissions are done independently on a per-cell basis. Specifically, each BS $i$ transmits $x_i=\sqrt{\SNR}\,s_i$ directly without compensating for inter-cell interference. The instantaneous SINR with optimal scheduling is therefore 
$$
\g{scp}(n)=\max_{1\leq k\leq n}\frac{|a_i(k)|^2}{1/\SNR+\beta^2|b_i(k)|^2}.
$$
In \cite{Sharif2005} it is shown that the cdf $F_\textsc{scp}$ of $\g{scp}:=\g{scp}(1)$ is
$$
F_\textsc{scp}(x)=1- \frac{e^{-x/\SNR}}{1+\beta^2 x}, \qquad x\geq 0.
$$ 
Hence, from the theory of order statistics we have that the cdf $F_{\textsc{scp}}^n$ of $\g{scp}(n)$ is 
\begin{displaymath}
F_{\textsc{scp}}^n(x)=\biggl(1- \frac{e^{-x/\SNR}}{1+\beta^2 x}\biggr)^n,\qquad x\geq 0.
\end{displaymath}
Having obtained the exact distribution we can now compute the mean SINR numerically. However, analytical solutions are hard to obtain and give little insight into the key quantities. Instead we will take an approach based on EVT in Section~\ref{sec:comparison}.

\section{Multicell Zero forcing beamforming}
We now consider ZFBF.  By definition of zero forcing there should be no interference for the scheduled users. It turns out that this can essentially be achieved with interference pre-subtraction. Specifically, let us assume we transmit
\begin{equation}
\label{eq:x_zf}
x_i=\SNR^{1/2}(1-|r_i|)^{1/2}s_i-r_i x_{i+1}
\end{equation}
where
\begin{displaymath}
	r_i:= \begin{cases}
				\beta\frac{b_i(k_i^*)}{a_i(k_i^*)}, & |a_i(k_i^*)|\geq \beta |b_i(k_i^*)| \\
				\beta\frac{b_i(k_i^*)}{a_i(k_i^*)}/ \bigl|\beta\frac{b_i(k_i^*)}{a_i(k_i^*)}\bigr|, &  |a_i(k_i^*)|< \beta |b_i(k_i^*)| 
				\end{cases}
\end{displaymath}
for all cells $i$. By solving \eqref{eq:x_zf} as a difference equation we obtain the coefficients of the beamforming matrix B,
\begin{displaymath}
	[B]_{i,j}=\begin{cases}
	0, & i>j\\
	(1-|r_j|^2)^{1/2}, &i=j\\
	(1-|r_j|^2)^{1/2}\prod_{l=i}^{j-1}{r_l}, &i<j.
	\end{cases}
\end{displaymath}
From \eqref{eq:x_zf} we can deduce directly that the per-cell power constraint $\E|x_i|^2\leq \SNR$ is satisfied since $\E(x_{i+1}s_i^*)=0$. Furthermore, if  $|a_i(k_i^*)|\geq \beta |b_i(k_i^*)|$ then
\begin{equation}
	y_i(k_i^*)=\bigl(|a_i(k_i^*)|^2-\beta^2|b_i(k^*_i)|^2\bigr)^{1/2}s_i+z_i(k_i^*).
\label{eq:channel_zf}
\end{equation}
Thus, the interference is eliminated at the expense of a power penalty. 

\subsection{Scheduling}
In order to characterize the performance of ZFBF we need to specify a particular scheduling policy.
From \eqref{eq:channel_zf} we can immediately conclude that optimal scheduling amounts to
$$k_i^*=\argmax_{1\leq k\leq n} |a_i(k_i)|^2-\beta^2|b_i(k_i)|^2.$$
The instantaneous post-scheduling SINR is now 
$$\g{zf}(n)=\max_{1\leq k\leq n}  \SNR\bigr[ |a_i(k)|^2-\beta^2|b_i(k)|^2\bigr]_+,$$
where $[\,\cdot\,]_+:=\max\{\,\cdot\,,0\}$.
Note that it is the received signal power after interference cancellation that determines the final performance. 
In the Appendix we find that the cdf of $\g{zf}:=\g{zf}(1)$ is 
\begin{equation}\label{eq:FZFBF}
F_{\textsc{zf}}(x)=1- \frac{e^{-x/\SNR}}{1+\beta^2 },\qquad x\geq 0.
\end{equation}
Hence, the cdf $F_{\textsc{zf}}^n$ of $\g{zf}(n)$ is
\begin{displaymath}
F_{\textsc{zf}}^n(x)=\biggl(1- \frac{e^{-x/\SNR}}{1+\beta^2 }\biggr)^n,\qquad x\geq 0.
\end{displaymath}

We also consider two suboptimal scheduling policies that have previously been proposed in the literature~\cite{Somekh2009,Jing2008}. The first policy is to schedule the user with the largest gain to the ``local'' BS,
\begin{displaymath}
	k_i^*=\argmax_{1\leq k\leq n} |a_i(k_i^*)|^2. 
\end{displaymath}
To denote the resulting instantaneous SINR we use $\g{zf,2}(n)$. 
The second policy is to schedule the user with largest ratio between the gains to ``local'' BS the ``non-local'' BS,  
\begin{displaymath}
	k_i^*=\argmax_{1\leq k\leq n} \frac{|a_i(k_i^*)|^2}{|b_i(k_i^*)|^2}. 
\end{displaymath}
In line with the previous notation we use $\g{zf,3}(n)$ to denote the resulting instantaneous SINR.

\section{Asymptotic results for the mean SINR}
\label{sec:comparison}
In this section we obtain some asymptotic results on the performance of ZFBF and SCP. We first note that $\g{scn}(n)$, $\g{scp}(n)$ and $\g{zf}(n)$ can all be viewed as the largest order statistics from a sample of size $n$. Based on this observation we make use of Extreme Value Theory (EVT)~\cite{Galambos1987,deHaan2006}, which is concerned with the asymptotic distribution of the largest order statistics. 

In the sequel, it will be convenient to extend the definitions of $\g{scn}(y),\g{scp}(y),\g{zf}(y)$ and $\g{zf,2}(y)$ to all $y\in\mathbb{R}_+$. To this end we take the distributions $F_\textsc{scn}^y,F_\textsc{scp}^y$ and $F_\textsc{scn}^y$ as definitions of $\g{scn}(y),\g{scp}(y)$ and $\g{zf}(y)$ for non-integers $y$. 


\subsection{Some Extreme Value Theory}
\label{sec:AsymptoticResults}
It is readily shown that $\Gamma_{\chi}$, $\chi\in\{\textsc{scn},\textsc{scp},\textsc{zf}\}$,
are all in the domain of attraction of the Gumbel distribution (see the Appendix for technical conditions). Thus, according to EVT there exist normalizing functions $\mu_\chi(y)$ and $\nu_\chi(y)$ such that
\begin{equation}
\label{eq:convergence}
\lim_{y\to \infty} F^y_\chi\bigl(\mu_{\chi}(y) + \nu_{\chi}(y)x\bigr)=G(x)\quad\text{for all $x$},
\end{equation}
where $G(x):=e^{-e^{-x}}$ is the Gumbel distribution. Furthermore, the normalizing functions can be selected to be
\begin{equation}
\label{eq:normalizing_functions}
\mu_{\chi}(y)= g_{\chi}(y)\quad\text{and}\quad\nu_{\chi}(y)= g_{\chi}(ye)-g_{\chi}(y),
\end{equation}
where $g_{\chi}(y):=F_{\chi}^{-1}(1-1/y)$.

The relationship in \eqref{eq:convergence} corresponds to convergence in distribution. Additionally, one can also show that there is convergence in moments~\cite{Song2006}. This means that we once we obtain the normalizing functions we also have a characterization of the asymptotic behavior of the mean. In particular, by computing the first moment of the Gumbel distribution we get
\begin{equation*}
\overline{\Gamma}_\chi(n):=\E\,\Gamma_{\chi}(n)\approx \mu_{\chi}(n)+\gamma \nu_{\chi}(n),
\end{equation*}
for large number of users $n$. 

\subsection{Explicit relationships for the normalizing functions}
For $\g{scn}$ and $\g{zf}$ it is straightforward to find the normalizing functions from \eqref{eq:normalizing_functions}. In particular, we have 
\begin{align}
\mu_{\textsc{scn}}(y)&=\SNR\ln y \label{eq:mu_scn}\\
\mu_{\textsc{zf}}(y)&=\SNR\ln y -\SNR\ln(1+\beta^2)\label{eq:mu_zf} \\
\nu_{\textsc{scn}}(y)&=\nu_{\textsc{zf}}(y)=\SNR. \notag
\end{align} 
Unfortunately, for  $\g{scp}$ the normalizing functions can not be expressed in terms of elementary functions. To proceed we make use of the Lambert $W$ function which is defined through the relation $W(x)e^{W(x)}=x$~\cite{Horfar2008}.  We then obtain
\begin{align*}
\mu_{\textsc{scp}}(y)&=\SNR W\biggl(\frac{y}{\beta^2\SNR}e^{\frac{1}{\beta^2\SNR}}\biggr)- \frac{1}{\beta^2},\\
\nu_{\textsc{scp}}(y)&=\SNR W\biggl(\frac{ye}{\beta^2\SNR}e^{\frac{1}{\beta^2\SNR}}\biggr) -\SNR W\biggl(\frac{y}{\beta^2\SNR}e^{\frac{1}{\beta^2\SNR}}\biggr)\underset{y\rightarrow\infty}{\longrightarrow} \SNR,
\end{align*}
where the limit can be inferred from  $W(x)=\ln x- \ln\ln x+O(\frac{\ln\ln x}{\ln x})$~\cite{Horfar2008}.
To gain more insight into the limiting behavior one can use more refined asymptotic expansions of $W(x)$. However, we will focus next on an an alternative indirect characterization of $\mu_{\textsc{scp}}(y)$.

\subsection{Implicit relationships for the normalizing functions}
 Interestingly, we can express  $\mu_{\textsc{scp}}(y)$ and $\mu_{\textsc{zf}}(y)$ implicitly in terms of $\mu_{\textsc{scn}}(y)$. From \eqref{eq:mu_scn} and \eqref{eq:mu_zf} we see that  $$\mu_{\textsc{zf}}\bigl(y(1+\beta^2)\bigr)=\mu_{\textsc{scn}}(y).$$
Similarly, from the observation
\begin{equation*}
1-F_{\textsc{scp}}(\mu_{\textsc{scn}}(y))=\frac{1}{y(1+\beta^2\SNR\ln y)}
\end{equation*}
we obtain the following relationship
\begin{equation*}
\label{eq:u}
\mu_{\textsc{scp}}\bigl(y(1+\beta^2\SNR\ln y)\bigr)=\mu_{\textsc{scn}}(y).
\end{equation*}
All in all we can infer from above that 
\begin{equation}\label{eq:SINRapprox}
\G{scp}\bigl(n(1+\beta^2\SNR \ln n)\bigr)\approx\G{scn}(n)\approx\G{zf}\bigl(n(1+\beta^2)\bigr),
\end{equation}
for large number of users $n$. 
Thus,  to attain the same mean SINR as in a single-cell network with $n$ users one needs asymptotically $n(1+\beta^2\SNR\ln n)$ users per cell with SCP and $n(1+\beta^2)$ users per cell with ZFBF.
It is interesting to note that ratio of required users with SCP to ZFBF is not bounded, but grows logarithmically with the number of users $n$. We also point out that the ratio is linear in $\SNR$. Thus, ZFBF is increasingly beneficial with increasing SNRs which is consistent with common knowledge. 

\section{Equalities and bounds for the mean SINR}
Even though the above analysis reveals the asymptotic behavior of the mean SINRs it fails to say anything about the rates of convergence. Furthermore, EVT is not directly applicable to the study of $\g{zf,2}(n)$ and $\g{zf,3}(n)$ since they can not be formulated as order statistics. Below we give some exact result together with several upper and lower bounds. The proofs can be found in the Appendix. We will assume in the following that $\g{scn}$, $\g{scp}$ and $\g{zf}$ are not identical, i.e. $\beta\neq 0$.
 
We first consider some results pertaining to ZFBF and suboptimal scheduling. 
\begin{proposition}\label{proposition1}
Let the user $k$ with the largest ratio $|a_i(k)|^2/|b_i(k)|^2$ be scheduled in each cell $i$. The mean SINR with ZFBF has the following upper bound
\begin{equation*}
\G{zf,3}(n)<2\SNR.
\end{equation*}
\end{proposition}

Proposition \ref{proposition1} is interesting because the upper bound is independent of the number of users per-cell. Clearly, the benefit of adding more users is severely limited. This is in contrast with the other suboptimal scheduling strategy which we consider below.
\begin{proposition}
Let the user $k$ with the largest gain $|a_i(k)|^2$ be scheduled in each cell $i$. The mean SINR with ZFBF is
\begin{align}
\G{zf,2}(n)&=\SNR H_n-\SNR\beta^2\Bigl(1- n B\Bigl(\tfrac{1+\beta^2}{\beta^2},n\Bigr)\Bigr)\notag\\
					 &\leq\SNR H_n-\SNR\beta^2\frac{n}{n+1},\label{zf2eq}
\end{align}
where $B(x,y)$ denotes the beta function~\cite{Gradshteyn}. The inequality is strict for all $0<\beta^2<1$. 
\end{proposition}

From \eqref{zf2eq} and the asymptotic expansion $H_n\sim\ln n+\euler$ it follows that
$$\G{zf,2}\bigl(ne^{\beta^2}\bigr)\approx\ln n+\euler\approx\G{scn}(n).$$
for $n$ large. Thus, compared to optimal scheduling we need approximately $35\%$ more users to attain the same mean SINR when $\beta^2=1$.
We next give an explicit expression for the mean SINR with optimal scheduling.
\begin{proposition}
The mean SINR with ZFBF and optimal scheduling is
\begin{align}
\G{zf}(n)&=\SNR H_n- \SNR\sum_{k=1}^n\biggl(\dfrac{\beta^2}{1+\beta^2}\biggr)^k\dfrac{1}{k}\label{zfeq}\\
					 &>\SNR H_n-\SNR\ln(1+\beta^2),\notag
\end{align}
where the last inequality is asymptotically tight. Additionally,
\begin{equation}\label{zfineq}
\G{zf}\bigl(n(1+\beta^2))<\G{scn}(n)<\G{zf}\bigl(n(1+\tfrac{n+1}{n}\beta^2)).
\end{equation}
\end{proposition}

We next give an upper bound to the performance of SCP with optimal scheduling.
\begin{proposition}\label{proposition2}
Assume SCP and optimal scheduling. The mean SINR satisfies the following upper bound
\begin{equation}\label{scpineq}
\G{scp}\bigl(n(1+\beta^2\SNR \ln n)\bigr)<\G{scn}(n).
\end{equation}
\end{proposition}

Note that we already know from Section~\ref{sec:comparison} that the inequality is asymptotically tight.
\section{Implications for the per-cell sum-rate}
We now briefly consider the per-cell sum-rates. Define
\begin{displaymath}
	C_{\chi}(n):=\E\log_2\bigl(1+\Gamma_{\chi}(n)\bigr)
\end{displaymath}
for $\chi=\{\textsc{scn},\textsc{scp},\textsc{zf}\}$. Unfortunately,  the concavity of the $\log_2\bigl(1+(\cdot)\bigr)$ function prevents most of the results concerning the mean SINR do not automatically carry over to the per-cell sum-rate. However, we still have the following results. 
\begin{proposition}\label{prop:sr}
The per-cell sum-rate with SCP and optimal scheduling satisfies the following bounds
$$\log_2(1+\SNR\ln n)< {C}_{\textsc{scp}}\bigl(n(1+\beta^2\SNR\ln n)\bigr)<\log_2(1+\SNR H_n)$$
The per-cell sum-rate with ZFBF and optimal scheduling satisfies
$$\log_2(1+\SNR\ln n)< {C}_{\textsc{zf}}\bigl(n(1+\beta^2)\bigr)<\log_2(1+\SNR H_n),$$	
for $n$ sufficiently large.
\end{proposition}

The above results together with \eqref{eq:SINRapprox} suggest the approximation 
\begin{equation}\label{eq:RateApprox}
C_\textsc{scp}\bigl(n(1+\beta^2\SNR \ln n)\bigr)\approx C_\textsc{scn}(n)\approx C_\textsc{zf}\bigl(n(1+\beta^2)\bigr)
\end{equation}
for $n$ large. We will investigate the accuracy of the above relations in the next section. Proposition~\ref{prop:sr} also shows that the difference in the per-cell sum-rate with SCP and ZFBF goes to zero as the number of users goes to infinity. Let $\Delta C(n):=C_\textsc{zf}(n)-C_\textsc{scp}(n)$ and consider the estimate
\begin{equation}\label{convergence}
\begin{split}
\Delta C(n)&\approx\log_2\bigl(1+\mu_\textsc{zf}(n)\bigr)-\log_2\bigl(1+\mu_\textsc{scp}(n)\bigr)\\
		&=\log_2\biggl(1+\frac{\ln(1+\beta^2\SNR\ln t)-\ln(1+\beta^2)}{1/\SNR+\ln t}\biggr)\\
		&\approx\log_2(e)\frac{\ln\bigl(\tfrac{\beta^2}{1+\beta^2}\SNR\ln t\bigr)}{\ln t}
\end{split}
\end{equation}
where $t$ is the unique solution to $n=t(1+\beta^2\SNR\ln t)$. %
Hence $\Delta C(n)$ goes to zero, but the convergence is extremely slow. 
\section{Numerical results}
\label{sec:Sim}
\begin{figure}[t!]
\centering
\includegraphics[width=.7\linewidth]{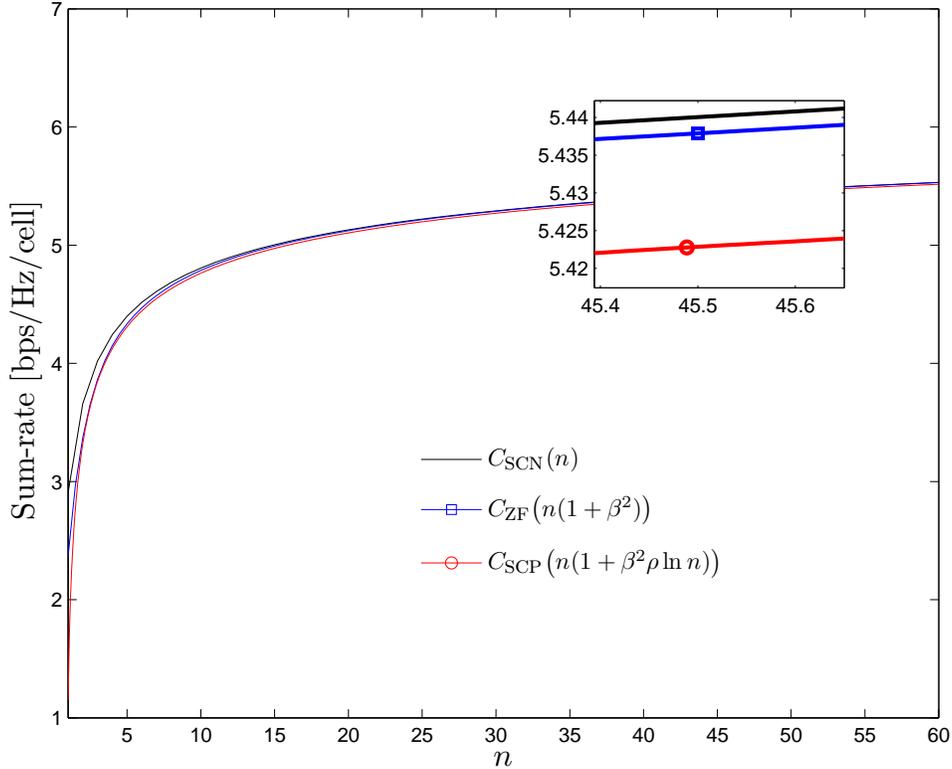}
\caption{The per-cell sum-rate for $\beta=0$ (upper bound) and $n$ users, ZFBF and $n(1+\beta^2)$ users, and SCP and $n(1+\beta^2\SNR\ln n)$ users.}
\label{fig:RateApprox}
\end{figure}

In this section we illustrate some our results through Monte Carlo simulations. We first consider the approximate relationship in \eqref{eq:RateApprox}. Specifically, in Fig.~\ref{fig:RateApprox} we plot the sum-rate per-cell corresponding to
\begin{itemize}
\item[(i)] a SCN scenario with $n$ users, 
\item[(ii)] ZFBF with $n(1+\beta^2)$ users per-cell and 
\item[(iii)] SCP with $n(1+\beta^2\SNR\ln n)$ users per-cell 
\end{itemize}
in the same plot. In all three cases the mean SNR is $\SNR=10$~dB and for (ii) and (iii) we have $\beta=1$. Observe that there is a remarkably good fit between the three graphs even for small $n$. Thus, the approximations in \eqref{eq:RateApprox} seems to be well justified.  The magnified section of the plot also reveals that the ordering between (i) and (ii) is as expected from \eqref{zfineq}. However, we point of that part of the difference is likely to result from the concavity of the rate function. The ordering of (i) and (iii) is also as one would expect from \eqref{scpineq}. However, in this case the concavity of the rate function is likely to lead to a small decrease in the difference as one would otherwise expect. 

The large difference in the number of users per cell between multicell ZFBF and SCP to attain the same rate is also interesting. To exemplify one needs over $240$ users with SCP as opposed to approximately $20$ users with ZFBF to attain the same rate as with a SCN and $10$ users.

Next we plot the sum-rate per-cell corresponding to a SCN, multicell ZFBF and SCP for the same number of users. Note that there is a significant gain with ZFBF over SCP. In accordance with \eqref{eq:convergence} there is little reduction in the gain even for very large number of users. The convergence of the two curves appears to have little impact in the pre-asymptotic user regime.
\begin{figure}[t!]
\centering
\includegraphics[width=.7\linewidth]{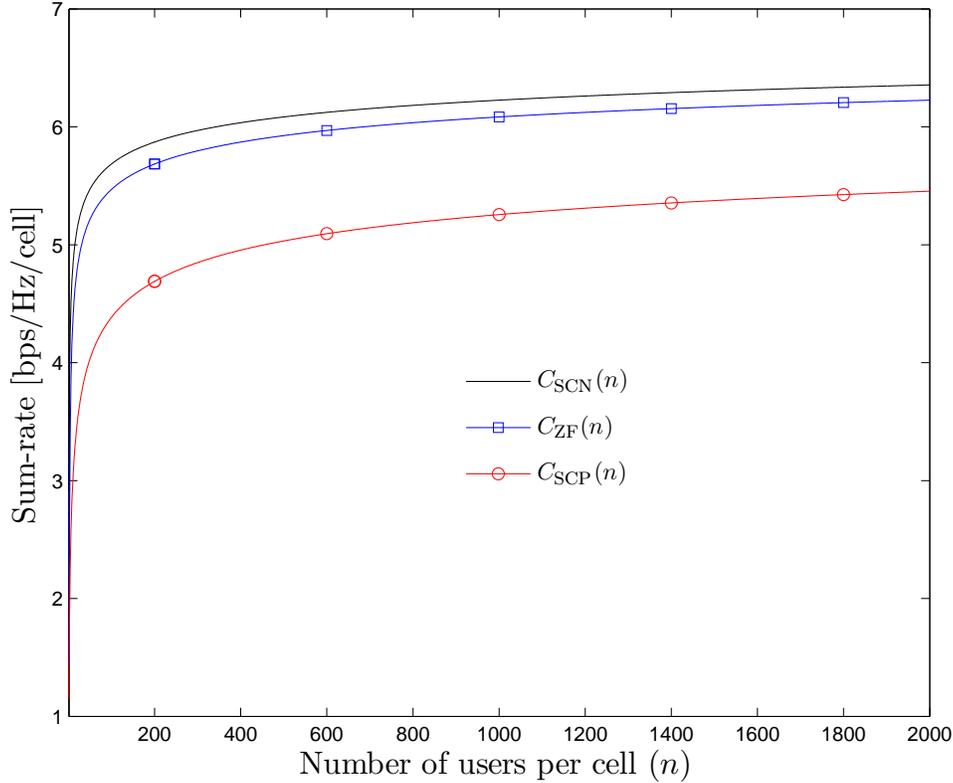}
\caption{The per-cell sum-rate for $\beta=0$ (upper bound), ZFBF and SCP as a function of the number of users per cell $n$.}
\label{fig:RateVsUsers}
\end{figure}

\section{Conclusion}
We have considered coordinated multicell ZBFB on the fading downlink of linear cell-array. The beamforming coefficients and the optimal scheduling policy under a per-base power constraint were both identified. Furthermore, the resulting mean post-scheduling SINR  was extensively studied. To put the performance in perspective SCP with optimal scheduling was used as a benchmark. Specifically, we gave asymptotic expressions for the additional number of users per cell to compensate for inter-cell interference with ZFBF and SCP. The difference in per-cell sum-rate between SCP and multicell ZFBF goes to zero as the number of users goes to infinity. However, we demonstrated that the convergence is too slow to have any practical impact. 

\section*{Appendix}

\subsection{$\g{scn},\g{scp}$ and $\g{zf}$ are in  the domain of the Gumbel distribution}
The claim follows immediately from the following result due to Von Mises\cite{Balkema1972}:\\
Suppose $X$ is random variable with cdf $F(x)$ and a pdf $f(x)$ which is positive and differentiable on a neighborhood of $x^*:=\sup\{x|F(x)<1\}$. If
\begin{equation}
\lim_{x\rightarrow x^*}\dfrac{\mathrm{d}}{\mathrm{d}x}\biggl(\frac{1-F(x)}{f(x)}\biggr)=0,
\end{equation}
then $X$ is in the domain of attraction of the Gumbel distribution.

\subsection{The distribution of $\g{zf}(n)$ is given according to \eqref{eq:FZFBF}}
We have by definition $\g{zf}\stackrel{d}{=}\SNR[|a_i(k)|^2-\beta^2|b_i(k)|^2]_+$ for a fixed $i$ and $k$. Since $\g{zf}$ cannot assume negative values we have $F_\textsc{zf}(x)=0$ for $x<0$. Let $F_\textsc{zf}(x|z)$ denote the cdf of $\g{zf}$ conditioned on $z=|b_i(k)|^2$, let $F_{|a|^2}(x)$ denote the cdf of $|a_i(k)|^2$ and let $f_{|b|^2}(x)$ denote the pdf of $|b_i(k)|^2$ . Note that $|a_i(k)|^2$ and $|b_i(k)|^2$ are exponential random variables with unit mean. By marginalizing over $|b_i(k)|^2$ the cdf of $\g{zf}$ can be expressed as
\ligning{
F_\textsc{zf}(x)&=\int\nolimits_0^\infty F_\textsc{zf}(x|z)f_{|b|^2}(z)\,\d z\\
             &=\int\nolimits_0^\infty F_{|a|^2}\Bigl(\tfrac{x+\beta^2z}{\SNR}\Bigr)f_{|b|^2}(z)\,\d z\\
             &=\int\nolimits_0^\infty \Bigl(1-e^{-(\frac{x}{\SNR}+\beta^2z)}\Bigr)e^{-z}\,\d z\\
             &=1-\frac{e^{-x/\SNR}}{1+\beta^2},
}
for $x>0$.


\subsection{Proof of Proposition $1$}
Let $A_k=|a_i(k)|^2,B_k=|b_i(k)|^2$ and $C_k:=A_k/B_k$ for a fixed $i$. 
We seek $\E\{C_{k^*}\}$ where $k^*=\argmax_{1\leq k\leq n}C_k$. The crucial point to observe is that knowing that $C_{k^*}$ is the largest out of $n$ variables do not give any extra information regarding $A_{k^*}$ once the exact value of $C_{k^*}$ is given. Thus,
$$f_{A_{k^*}}(x|C_{k^*}=z)=f_{A_k}(x|C_k=z)$$
for all $k$. Now since $A_k$ and $B_k$ have exponential distributions if follows that $C_k$ has a $F$-distribution~\cite[p. 946]{Abramowitz} with pdf
$$f_{C_k}(z)=\frac{1}{(1+z)^2},\quad z\geq 0.$$
Furthermore, $C_k$ conditioned on $A_k$ has an inverse exponential distribution with pdf
$$f_{C_k}(z|A_k=x)=\frac{x}{z^2}e^{-x/z},\quad z\geq 0.$$
Based on Bayes' theorem we now obtain
\begin{align*}
f_{A_{k^*}}(x|C_{k^*}=z)&=\frac{f_{A_k}(x)f_{C_k}(z|A_k=x)}{f_{C_k}(z)}\\
										&=\Bigl(1+ \frac{1}{z}\Bigr)^2x\,e^{-\bigl(1+\frac{1}{z}\bigr)x}.
\end{align*}
This is a Gamma distribution~\cite[p. 103]{Papoulis} with mean
$$\E\{A_{k^*}|C_{k^*}=z\}=\frac{2}{\bigl(1+\frac{1}{z}\bigr)^2}<2.$$
Thus, regardless of the distribution of $C_{k^*}$ we have $\E\{A_{k^*}\}< 2$. Finally,
$$\G{zf,3}(n)=\SNR\E[A_{k^*}-\beta^2 B_{k^*}]_+<2\SNR $$
which is the desired result.

\subsection{Proof of Proposition $2$}
Throughout the proof of Proposition $2$ we let $\SNR=1$ for simplicity. However, the general results follow by noting that the SINR is linear in $\SNR$ for ZFBF.

Let $A_k:=|a_i(k)|^2$, $B_k:=\beta^2|b_i(k)|^2$ and $k^*:=\argmax_{1\leq k\leq n} A_k$. Since $A_k$ and $B_k$ are exponential random variables it follows that $A_{k^*}$ has pdf
$$
f_{A_{k^*}}(x)=ne^{-x}\bigl(1-e^{-x}\bigr)^{n-1},\quad x\geq 0
$$ 
and $B_{k^*}$ has pdf
$$
f_{B_{k^*}}(y)=\frac{1}{\beta^2}e^{-x/\beta^2},\quad x\geq 0.
$$
Now, define $B_{k^*}^{'}$ such that
$$
[A_{k^*}-B_{k^*}]_+=A_k-B_{k^*}^{'}.
$$
The distribution of $B_{k^*}^{'}$ conditioned on $A_{k^*}$ is then
$$
F_{B_{k^*}^{'}}(y|A_{k^*}=x)=\begin{cases}
1-e^{-y/\beta^{2}},&y\leq x\\
1,&y>x.
\end{cases}
$$ 
and the conditional mean is
\begin{align*}
\E\{B_{k^*}^{'}|A_{k^*}=x\}&=\int_0^{\infty}1-F_{B_{k^*}^{'}}(y|A_{k^*}=x)\mathrm{d}y\\
													 &=\beta^2\bigl(1-e^{-x}\bigr).
\end{align*}
Finally,
\begin{align*}
\G{zf,2}(n)&=\E\,[A_{k^*}-B_{k^*}]_+\\
           &=\iint_{x,y\geq 0} (x-y) f_{A_{k^*}}(x)f_{B_{k^*}^{'}}(y|A_{k^*}=x)\,\mathrm{d}y\mathrm{d}x\\
           &=\int_{x\geq 0}(x-\E\{B_{k^*}^{'}|A_{k^*}=x\})f_{A_{k^*}}(x)\,\mathrm{d}x\\
           &=\int_{x\geq 0}\bigl(x-\beta^2\bigl(1-e^{-x/\beta^2}\bigl)\bigl)f_{A_{k^*}}(x)\,\mathrm{d}x\\
           &=H_n-\int_{x\geq 0}\beta^2\bigl(1-e^{-x/\beta^2}\bigr)n e^{-x}\bigl(1-e^{-x}\bigr)^{n-1}\,\mathrm{d}x\\
           &=H_n-\beta^2+\beta^2n\int\nolimits_0^1 t^{1/\beta^2}(1-t)^{n-1}\,\d x\\
           &=H_n-\beta^2+\beta^2 n B(1+1/\beta^2,n)\\
           &\leq H_n-\beta^2+\beta^2\frac{1}{n+1}
\end{align*}
where use the substitution $t=1-e^{-x}$. The inequality follows from observing that Beta-function is monotonically decreasing in both variables. Thus $B(1+1/\beta^2,n)\leq B(2,n)=\frac{1}{n(n+1)}$ with equality only for $\beta^2=1$.

Before we prove Proposition $3$ we state the following useful result on the harmonic numbers. 

\subsection{Result on the harmonic numbers}
Let $x\geq1$, the harmonic numbers satisfy the following relations
\begin{align}
H_x&=\ln x + \euler+\epsilon(x) \label{eq:epsilon}\\
	 &=\ln x + \euler+\frac{1}{2x}-\eta(x),\label{eq:eta}
\end{align}
where $\epsilon(x)$ and $\eta(x)$ are positive, monotonically decreasing functions~\cite{Chen2003}.

\subsection{Proof of Proposition $3$}
\subsubsection{Proof of \eqref{zfeq}}
A direct calculation gives
\begin{align*}
\G{zf}(n)&=\int\nolimits_0^\infty 1 - F^n_\textsc{zf}(x)\,\d x\\
												    &=\int\nolimits_0^\infty 1 - \biggl(1-\frac{e^{-x/\SNR}}{1+\beta^2}\biggr)^n \,\d x\\
												    &=\SNR\int\nolimits_{\frac{\beta^2}{1+\beta^2}}^1 \frac{1-z^n}{1-z}\,\d z\\
												    &=\SNR\int\nolimits_{\frac{\beta^2}{1+\beta^2}}^1 \sum_{k=1}^n z^{k-1}\,\d z\\
												    &=\SNR\sum_{k=1}^n\frac{1}{k}-\SNR\sum_{k=1}^n \Biggl(\frac{\beta^2}{1+\beta^2}\Biggr)^k\frac{1}{k}\\
												    &>\SNR H_n-\SNR\ln(1+\beta^2)
\end{align*}
where we have used the substitution $z=1-\frac{e^{-x}}{1+\beta^2}$. The inequality follows from the identity~\cite[p. 68]{Abramowitz} $$\ln(x)=\sum_{k=1}^\infty\left(\frac{x-1}{x}\right)^k\frac{1}{k}.$$

\subsubsection{Proof of \eqref{zfineq}}
The left side follows from the following calculation 
\ligning{
\G{zf}\bigl(n(1+\beta^2)\bigr)&=\int\nolimits_0^\infty 1 - \biggl(1-\frac{e^{-x/\SNR}}{1+\beta^2}\biggr)^{n(1+\beta^2)} \,\d x\\
													    &<\int\nolimits_0^\infty 1 - \bigl(1-e^{-x/\SNR}\bigr)^n \,\d x\\
														  &=\G{scn}(n)}
where we use Bernoulli's inequality, $(1+x)^r>1+rx$ for $x>-1$ and $r>1$~\cite{Mitrinovic1972}.

We now turn to the right hand side of the inequality. Let $y:=n(1+\frac{n+1}{n}\beta^2)$. From \eqref{eq:eta} we have
\ligning{
\G{zf}(y)/\SNR	 &>\ln y +\euler +\frac{1}{2y} -\eta(y)-\ln\bigl(1+\beta^2\bigr)\\
					 &=\ln n +\euler+\frac{1}{2y}-\eta(y)+\ln\Bigr(1+\frac{1}{n}\frac{\beta^2}{1+\beta^2}\Bigr)}
					 and
					 \ligning{
\G{scn}(n)/\SNR &=\ln n + \euler +\frac{1}{2n}-\eta(n).}
Thus, since $\eta(x)$ is monotonically decreasing it is sufficient to show
\begin{equation}\label{eq:ineq}
\ln\Bigr(1+\frac{1}{n}\frac{\beta^2}{1+\beta^2}\Bigr)+\frac{1}{2n(1+\beta^2+\frac{1}{n}\beta^2)}\geq\frac{1}{2n}.
\end{equation}
To proceed we use the following inequality~\cite[p. 68]{Abramowitz}
$$\ln\bigg(1+\frac{1}{x}\biggr)>\frac{1}{x+1},\quad x>0.$$
Applied to the left side of~\eqref{eq:ineq} this gives
$$\frac{\beta^2}{n(1+\beta^2)+\beta^2}+\frac{1}{2n(1+\beta^2+\frac{1}{n}\beta^2)}=\frac{1+2\beta^2}{1+\beta^2+\frac{1}{n}\beta^2}\frac{1}{2n}.$$
Thus, $\G{zf}\left(n(1+\frac{n+1}{n}\beta^2)\right)>\G{scn}(n)$ for $n\geq 1$.

Before we prove Propostion 4 we will review the probability integral transform theorem~\cite{Angus1994}.

\subsection{The probability integral transform theorem}
Suppose $X$ is a random variable with continuous cdf $F_X$. By the integral transform theorem we have that $U:=F_X(X)$ is a uniform random variable on $[0,1]$. The following extension is straight forward. Assume $F_X(0)=c$ and define $X_+:=[X]_+$. The cdf of $X_+$ is then $F_{X_+}(x)=\max\{F(x),c\}:=[F(x)]_c$. Thus,
$$F_{X_+}(X_+)=[F_{X}(X_+)]_c=[F_X(X)]_c=[U]_c.$$
Furthermore,
$$X_+=F_{X_+}^{-1}\left([U]_c\right)=F_{X_+}^{-1}\left(U\right).$$

\subsection{Proof of Proposition $4$}
To prove \eqref{scpineq} the following results will be convenient.
\begin{align}
&\g{scn}(y)\stackrel{d}{=}\g{scp}(y)+\SNR\ln\big(1+\beta^2\g{scp}(y)\bigr) \label{ass1}\\
&F_U\bigl(\E\,U^{1/y}\bigr)=1-\frac{1}{y+1}>1-\frac{1}{y} \label{ass2}\\
&\G{scp}(y)>\SNR\ln n \label{ass3}\\
&\E\ln\bigl(1+\beta^2\g{scp}(y)\bigr)>\ln\bigl(1+\beta^2\SNR\ln n \bigr) \label{ass4}
\end{align}
Here $U$ is uniformly distributed on $[0,1]$ and $n$ is the unique solution to $y=n(1+\beta^2\SNR\ln n)\geq 1$.  Assuming the above results to be true, we obtain
\ligning{
\G{scp}(y)&=\G{scn}(y)-\SNR\E\ln\big(1+\beta^2\g{scp}(y) \bigr)\\
												&<\SNR\ln y +\SNR\euler+\SNR\epsilon(y)-\SNR\ln\big(1+\beta^2\SNR\ln n \bigr)\\
												&=\SNR\ln n +\SNR\euler+\SNR\epsilon(y)\\
												&<\SNR\ln n +\SNR\euler+\SNR\epsilon(n)\\
												&=\G{scn}(n).
}
which is the desired result. The last inequality follows follows from the fact that $\epsilon(x)$ is monotonically decreasing.

\subsubsection{Proof of \eqref{ass1}}
By the probability integral transform theorem we have 
\ligning{
U\stackrel{d}{=}F^y_\textsc{scn}\Bigl(\g{scn}(y)\Bigr)
\stackrel{d}{=}F^y_\textsc{scp}\Bigl(\g{scp}(y)\Bigr).
}
This in turn yields
\ligning{
\g{scn}(y)&\stackrel{d}{=}[F^y_\textsc{scn}]^{-1}\circ F^y_\textsc{scp}\Bigl(\g{scp}(y)\Bigr)\\
											 &=-\SNR\ln\biggl(1-\Bigl[F^y_\textsc{scp}\bigl(\g{scp}(y)\bigr)\Bigr]^{1/n}\biggr)\\
											 &=-\SNR\ln\biggl(\frac{e^{-\g{scb}(y)/\SNR}}{1+\beta^2\g{scb}(y)}\biggr)\\
											 &=\g{scb}(y)+\SNR\ln\bigl(1+\beta^2\g{scb}(y)\bigr).
}

\subsubsection{Proof of \eqref{ass2}}
The pdf and cdf of $U$ are
$F_U(x)=x$, $f_U(x)=1$, $0\leq x\leq 1$. Thus, 
\ligning{F_U\bigl(\E\,U^{1/y}\bigr)=\E\,U^{1/y}=\int_0^1 f_U(x)x^{1/y}\d x=1-\frac{1}{y+1}.}

\subsubsection{Proof of \eqref{ass3}}
Applying the probability integral theorem we have $U\stackrel{d}{=}F^y_\textsc{scp}\bigl(\g{scp}(y)\bigr)$. Thus, $U^{1/y}\stackrel{d}{=}F_\textsc{scp}\Bigl(\g{scp}(y)\Bigr)$. Therefore, if $F_\textsc{scp}$ is concave we have
$$\E\,U^{1/n}\leq F_\textsc{scp}\Bigl(\G{scp}(y)\Bigr)$$ 
by Jensen's inequality. This in turn gives
\begin{equation}\label{eq:bongo}
\begin{split}
\G{scp}(y)&\geq F_\textsc{scp}^{-1}\bigl(\E\,U^{1/y}\bigr)
													>F_\textsc{scp}^{-1}\Bigl(1-\frac{1}{y}\Bigr)
													=\SNR\ln n
\end{split}
\end{equation}
where the second inequality follows from \eqref{ass3} and the last equality from the relation
$$F_\textsc{scp}(\SNR\ln n)=1-\frac{1}{n(1+\beta^2\ln n)}.$$
To prove the concavity of $F_\textsc{scn}$ we show that its second derivative is non-positive.  
\ligning{
\frac{\d^2}{\d x^2}\,F_\textsc{scn}(x)&=\bigl(1-e^{-g(x)}\bigr){''}\\
			                                   &=\Big(e^{-g(x)}g{'}(x)\Bigr){'}\\
			                                   &=-e^{-g(x)}\Bigl(\bigl(g{'}(x)\bigr)^2-g{''}(x)\Bigr)\\
			                                   &\leq 0
}
where $g(x):=x/\SNR+\ln(1+\beta^2 x)$.
\subsubsection{Proof of \eqref{ass4}}
Let $\Lambda(y):=\ln\Bigl(1+\beta^2\,\g{scp}(y)\Bigr)$. The cdf $F_{\Lambda}^y$ of $\Lambda(y)$ is then
\ligning{
F_{\Lambda}^y(x)&=F^y_\textsc{scn}\biggl(\frac{e^x-1}{\beta^2}\biggr)\\
					      &=\bigl(1-e^{-x+ \frac{e^x-1}{\SNR\beta^2}}\bigr)^y.
}
If $F_\Lambda:=F_{\Lambda}^1$ is concave we now have
\ligning{
\E\ln\bigl(1+\beta^2\g{scp}(y)\bigr)&=\E F_{\Lambda}^{-1}\bigl(U^{1/y}\bigr)\\
			                              &\geq  F_{\Lambda}^{-1}\bigl(\E\,U^{1/y}\bigr)\\
			                              &=\ln\Bigl(1+\beta^2 F_\textsc{scp}^{-1}\bigl(\E\,U^{1/y}\bigr)\Bigr)\\
			                              &>\ln\bigl(1+\beta^2\SNR\ln n \bigr),
}
where we use the probability integral transform theorem, Jensen's inequality and finally \eqref{ass3}. To prove the concavity of $F_{\Lambda}$ we demonstrate that its second derivative is non-positive.  
\ligning{
\frac{\d^2}{\d x^2}\,F_{\Lambda}(x)&=\bigl(1-e^{-g(x)}\bigr){''}\\
																&=-e^{-g(x)}\Bigl(\bigl(g{'}(x)\bigr)^2-g{''}(x)\Bigr)\\
																&=-e^{-g(x)}\Bigl(\Bigl(1 +\frac{e^x}{\SNR\beta^2}\Bigr)^2-\frac{e^x}{\SNR\beta^2}\Bigr)\\
																&< 0,
}
where $g(x):=x+\frac{e^x-1}{\SNR\beta^2}$.

\subsection{Proof of Proposition $5$}
From Jensen's inequality and Proposition $4$ we have 
\ligning{
C_\textsc{scp}\bigl(n(1+\beta^2\SNR\ln n)\bigr)&=\E\log_2\Bigl(1+\g{scp}\bigl(n(1+\beta^2\SNR\ln n)\bigr)\Bigr)\\
																	&<\log_2\Bigl(1+\E\,\g{scp}\bigl(n(1+\beta^2\SNR\ln n)\bigr)\Bigr)\\
																	&<\log_2\Bigl(1+\g{scn}(n)\Bigr)\\
																	&=\log_2\bigl(1+\SNR H_n\bigr)
}
Likewise, from Jensen's inequality and Proposition $3$ we have
\ligning{C_\textsc{zf}\bigl(n(1+\beta^2)\bigr)< \log_2\bigl(1+\SNR H_n\bigr).}
From \eqref{ass4} it immediately follows that
\ligning{C_\textsc{scp}\bigl(n(1+\beta^2\SNR\ln n)\bigr)> \log_2\bigl(1+\SNR\ln n\bigr).}

Finally we turn to the claim,
$$C_\textsc{zf}\bigl(n(1+\beta^2)\bigr)>\log_2(1+\SNR\ln n)$$
for $n$ sufficiently large. We first introduce the notation 
$$R(y):=\log_2\bigl(1+\g{zf}(y)\bigr)$$
and $R:=R(1)$. The cdf of $R$ is then $F_R(x)=F_\textsc{zf}\bigl(2^x-1\bigr).$ To prove the desired result we postulate a random variable $Z$ with cdf $F_Z$ such that $u(x):=F_Z^{-1}\circ F_{R}(x)$ is concave and
\begin{equation}\label{FE_y}
F_Z\bigl(\E\,Z(y)\bigr)> 1-\frac{1}{y}
\end{equation}
for $y$ sufficiently large. Here $Z(y)$ is defined through its cdf $F_{Z(y)}(x)=\bigl(F_Z(x)\bigr)^y$. By the integral transform theorem we then have 
$$R(y)\stackrel{d}{=}F_R^{-1}\circ F_Z\bigl(Z(y)\bigr)=u^{-1}\left(Z(y)\right)$$
where $u^{-1}(x)$ is convex since $u(x)$ is concave. The desired result then follows from Jensen's inequality since
\ligning{
C_\textsc{zf}\bigl(n(1+\beta^2)\bigr)&=\E R\bigl(n(1+\beta^2)\bigr)\\
																		 &\geq F_R^{-1}\circ F_Z\Bigl(\E\,Z\bigl(n(1+\beta^2)\bigr)\Bigr)\\
																		 &> F_R^{-1}\left(1- \frac{1}{n(1+\beta^2)}\right)\\
																		 &=\log_2\Bigl(1+F_\textsc{zf}^{-1}\Bigl(1- \frac{1}{n(1+\beta^2)}\Bigr)\Bigr)\\
																		 &=\log_2(1+\SNR\ln n).
}

To prove the existence of $Z$ we introduce the following quantities
\begin{align*}
h_1(x)&:= \tfrac{\beta^2}{1+\beta^2}+ \tfrac{1}{1+\beta^2}\tfrac{2^x-1}{\SNR}\\
x_m&:=  h_1^{-1}\Bigl(1-\tfrac{e^{-1}}{1+\beta^2}\Bigr)\\
c_2&:=h^{'}_1(x_m)\\
h_2(x)&:=1-\tfrac{e^{-1}}{1+\beta^2}+c_2(x-x_m)\\
x_e&:=h_2^{-1}(1).
\end{align*}
We now define $Z$ to have support $[0,x_e]$ and cdf
$$
F_Z(x):=\begin{cases}
h_1(x),& 0\leq x\leq x_m\\
h_2(x),& x_m< x\leq x_e.
\end{cases}
$$
Note that $F_Z$ has a continuous derivative on its support. To prove the concavity of $u(x)$ we fist show that the second derivative of $u(x)$ is negaive on $\Bigl[0,F_R^{-1}\Bigl(1-\frac{e^{-1}}{1+\beta^2}\Bigr)\Bigr)$ and then on $\Bigl(F_R^{-1}\Bigl(1-\frac{e^{-1}}{1+\beta^2}\Bigr),\infty]$. Since $u(x)$ has a continuous derivative it follows that $u(x)$ is concave on the whole of $[0,\infty)$.

For $x\in\Bigl[0,F_R^{-1}\Bigl(1-\frac{e^{-1}}{1+\beta^2}\Bigr)\Bigr)$ we have
\ligning{
u(x)&=\log_2\bigl(1+\SNR\bigl((1+\beta^2)F_R(x)-\beta^2\bigr)\bigr)\\
	  &=\log_2\Bigl(1+\SNR\Bigl(1-e^{-\frac{2^x-1}{\SNR}}\Bigr)\Bigr).
}
Now let $v(x)$ denote the argument of $\log_2(\cdot)$ above. By taking the second derivative of $u(x)$ we obtain
\ligning{
u''(x)&=\biggl(\frac{1}{\ln 2}\frac{v'(x)}{v(x)}\biggr)'\\
			&=\frac{1}{\ln 2}\frac{v''(x)}{v(x)}-\frac{1}{\ln 2}\frac{\bigl(v'(x)\bigr)^2}{v(x)^2}\\
			&=\frac{\ln 2\, 2^x e^{-\frac{2^x-1}{\SNR}}}{v(x)}\cdot\biggl\{1-\frac{2^xe^{-\frac{2^x-1}{\SNR}}}{1+\SNR\Bigl(1-e^{-\frac{2^x-1}{\SNR}}\Bigr)}-\frac{2^x}{\SNR}\biggr\}.
}
By applying the inequality
$e^{-x}\leq 1-x$ twice inside the curly brackets we get
\ligning{
u''(x)&\leq - \frac{\ln 2\, 2^x e^{-\frac{2^x-1}{\SNR}}}{v(x)} \frac{1}{\SNR}<0.
}

For $x\in\Bigl(F_R^{-1}\Bigl(1-\frac{e^{-1}}{1+\beta^2}\Bigr),\infty\Bigr)$ we have
$$u(x)=x_m+ \frac{1}{c_2}\bigl(F_R(x)+ \frac{e^{-1}}{1+\beta^2}-1\bigr).$$
By taking the second derivative we obtain
\ligning{
u''(x)&=\frac{1}{c_2}\biggl(1-\frac{e^{-\frac{2^x-1}{\SNR}}}{1+\beta^2}\biggr)''\\
			&=\frac{1}{c_2}\biggl(\frac{e^{-\frac{2^x-1}{\SNR}}}{1+\beta^2}\frac{2^x}{\SNR}\ln 2\biggr)'\\
			&=\frac{1}{c_2}\biggl(\frac{e^{-\frac{2^x-1}{\SNR}}}{1+\beta^2}\frac{2^x}{\SNR}(\ln 2)^2\biggr)\cdot\Bigl\{1- \frac{2^x}{\SNR}\Bigr\},
}
which is negative for $x>\log_2(\SNR)$. Hence $u{''}(x)$ is negative for $x>F_R^{-1}\Bigl(1-\frac{e^{-1}}{1+\beta^2}\Bigr)=\log_2(1+\SNR)$.

To prove \eqref{FE_y} we introduce the function
$$h_3(x):=\frac{\beta^2}{1+\beta^2}+ c_3x,$$
with $c_3:=\tfrac{1-e^{-1}}{(1+\beta^2)x_m}$. Note that $h_3(x)$ satisfies $h_3(x)>h_1(x)$ for $x\in(0,x_m)$. Hence,
\ligning{
\E\,Z(y)&=\int\nolimits_0^{x_e} 1-\bigl(F_Z(x)\bigr)^y\,\d x\\
				&=\int\nolimits_0^{x_m} 1-\bigl(h_1(x)\bigr)^y\,\d x+\int\nolimits_{x_m}^{x_e} 1-\bigl(h_2(x)\bigr)^y\,\d x\\
				&>\int\nolimits_0^{x_m} 1-\bigl(h_3(x)\bigr)^y\,\d x+\int\nolimits_{x_m}^{x_e} 1-\bigl(h_2(x)\bigr)^y\,\d x\\
				&=x_e- \frac{1/c_3}{y+1}\Bigl[\bigl(1-\tfrac{e^{-1}}{1+\beta^2}\bigr)^{y+1}-\bigl( \tfrac{\beta^2}{1+\beta^2}\bigr)^{y+1}\Bigr]\\
				&\qquad\qquad\qquad -\frac{1/c_2}{y+1}\Bigl[1-\bigl(1-\tfrac{e^{-1}}{1+\beta^2}\bigr)^{y+1}\Bigr].
}
Since $\E\,Z(y)$ goes to $x_e$ with increasing $y$ we have for $y$ sufficiently large
\ligning{
F_Z\bigl(\E\,Z(y)\bigr)&=1-\tfrac{e^{-1}}{1+\beta^2}+c_2\bigl(\E\,Z(y)-x_m\bigr).
}
Substituting with the lower bound for $\E\,Z(y)$ we obtain
$$F_Z\bigl(\E\,Z(y)\bigr)<1- \frac{\bigl(\frac{c_2}{c_3}-1)\bigl(1-e^{-1}\bigr)^{y+1}+1}{y+1}.$$
This completes the proof since
$$\frac{\bigl(\frac{c_2}{c_3}-1)\bigl(1-e^{-1}\bigr)^{y+1}+1}{y+1}<\frac{1}{y}$$
for $y$ sufficiently large.


\end{document}